\documentclass[twocolumn,showpacs,prl,aps,superscriptaddress]{revtex4}

\usepackage{amsmath}
\usepackage{amssymb}
\usepackage{graphicx}
\usepackage{upgreek}
\usepackage{bm}

\newcommand{\sprod}{\!\cdot\!}

\newcommand{\Tr}{\mathrm{tr}}

\newcommand{\rmi}{\mathrm{i}}
\newcommand{\rme}{\mathrm{e}}

\newcommand{\dyad}[1]{\mathbf{#1}}
\newcommand{\dGamma}{{\boldsymbol \Gamma}}

\newcommand{\vc}[1]{\mathbf{#1}}
\newcommand{\bd}[1]{{\boldsymbol #1}}

\newcommand{\jsum}{\sum_{j=0}^\infty \!{}^{{}^\prime}}
\newcommand{\msum}{\sum_{m=0}^\infty \!{}^{{}^\prime}}

\newcommand{\kB}{k_\mathrm{B}}

\newcommand{\Unr}{U^\mathrm{nr}_n}
\newcommand{\Ur}{U^\mathrm{r}_n}

\newcommand{\half}{{\textstyle\frac1{2}}}
\newcommand{\third}{{\textstyle\frac1{3}}}

\newcommand{\be}{\begin{equation}}
\newcommand{\ee}{\end{equation}}

\newcommand{\re}{\mathrm{Re}}

\begin{document}

\title{Temperature-independent Casimir--Polder forces in arbitrary
geometries}

\author{Simen {\AA}. Ellingsen}
\affiliation{Department of Energy and Process Engineering, Norwegian
University of Science and Technology, N-7491 Trondheim, Norway}
\email{simen.a.ellingsen@ntnu.no}
\author{Stefan Yoshi Buhmann}
\author{Stefan Scheel}
\affiliation{Quantum Optics and Laser Science, Blackett Laboratory,
Imperial College London, Prince Consort Road, London SW7 2AZ, United
Kingdom}

\date{\today}

\begin{abstract}
We show that the Casimir--Polder potential of a particle in an energy
eigenstate at nonretarded distance from a well-conducting body of
arbitrary shape is independent of the environment temperature. This is
true even when the thermal photon numbers at the relevant atomic
transition energies are large. A compact expression is obtained for
the temperature-independent potential, which can greatly simplify
calculations in nontrivial geometries for experimentally relevant
systems such as Rydberg atoms and polar molecules. We give criteria
for the validity of our temperature-independent result. They are
illustrated by numerical studies of a particle near a gold sphere or
inside a gold cylindrical cavity.
\end{abstract}

\maketitle

%%%%%%%%%%%%%%%%%%%%%%%%%%%%%%%%%%%%%%%%%%%%%%%%%%%%%%%%%%%%%%%%%%%%%%

\paragraph*{Introduction.}

Theoretical and experimental research into the Casimir and
Casimir--Polder (CP) effects \cite{casimir48} have seen a phenomenal
surge of activity over the last good decade \cite{lecturenotes}.
These effects, together with the van der Waals forces often
collectively termed dispersion forces, are consequences of ever
present fluctuations, quantum and thermal, of charges, currents and
electromagnetic fields.

Motivated by experimental progress and an increasing number of
technological applications, one has arrived at a number of very
general insights into the behaviour of dispersion forces in recent
years. It was shown that the Casimir interaction between bodies that
are mirror images of each other is always attractive \cite{kenneth06}.
A recently proven Earnshaw-type theorem generalises this result,
demonstrating that Casimir forces between dielectric bodies of
arbitrary shapes in vacuum do not support stable equilibria
\cite{rahi10}. General duality symmetry relations relate dispersion
forces between electric vs magnetic objects \cite{buhmann09}, while
scaling relations constrain their dependence on body separations and
sizes \cite{buhmann10}.

With experiments typically being conducted at room temperature, the
impact of thermal photons on the CP potential \cite{henkel02} has been
of particular practical interest. It has recently been studied for
various non-equilibrium scenarios \cite{antezza05,ducloy06,buhmann08}
where repulsive or spatially oscillating (transient) forces
have been predicted \cite{ellingsen09}. On the other hand, it
has also been shown that the CP potential of particles at nonretarded
separations (small compared to the transition wavelengths) from a
metal plate is independent of the ambient temperature, although the
available thermal photon numbers at the relevant transition energies
can be very large \cite{ellingsen10}.

Because the geometry and temperature dependencies of dispersion
forces are closely intertwined \cite{weber10}, one might suspect this
temperature-invariance to be an artefact of the planar geometry.Quite
the contrary, we will show in this Letter that this result extends to
particles close to conducting bodies of arbitrary shape. The derived
temperature-independent potential is governed by the electrostatic
Green tensor. For geometries tractable via an image-charge method, it
is simply the quantum-averaged interaction of the atomic dipole with
its images as produced by the curved body surfaces.

This very simple solution to an initially complex problem of
nontrivial geometry out of thermal equilibrium could greatly simplify
technologically and experimentally important calculations involving
such experimentally promising systems. It shows that cooling is not
able to reduce or alter the nonretarded CP potential of a particle
near any metal body. Two important systems for which this
temperature-invariance holds in typical experimental set-ups are
ground state polar molecules \cite{ellingsen09} and atoms in highly
excited Rydberg states \cite{crosse10}. For these particles, the
nonretarded regime extends to several micrometers (polar molecules)
and several hundred micrometers (Rydberg atoms), respectively.

We begin by proving that the temperature-dependence of the CP
force is cancelled at nonretarded separations for bodies of arbitrary
shapes and give criteria for the validity of the
temperature-independent result. These are illustrated by two numerical
calculations for particles in nontrivial geometries.

%%%%%%%%%%%%% SECTION %%%%%%%%%%%%%%%%%%%%%%%%%%%%%%%%%%%%%%%%%%%%%%%%

\paragraph*{Temperature-independence in the nonretarded regime.}

In this section we will show the following: Consider a particle in the
vicinity of metallic bodies of arbitrary shape, such that all relevant
particle--body separations are much smaller than the dominant
intra-atomic transition wavelengths (nonretarded regime). Assuming
further that the reflectivities of the bodies at frequencies up to the
atomic transition frequencies do not differ from the static
reflectivies, the CP interaction is independent of temperature.

As shown in Ref.~\cite{buhmann08}, the CP potential of a particle
prepared in an energy eigenstate $|n\rangle$ may be given as
$U_n(\vc{r})=\Unr(\vc{r})+\Ur(\vc{r})$ with nonresonant and
resonant contributions
\begin{eqnarray}
  &&\Unr(\vc{r}) = -
\frac{\kB T}{\varepsilon_0}
\jsum\mathrm{Tr}[\bm{\alpha}_n(\rmi\xi_j)\sprod
 \dGamma_{\rmi\xi_j}(\vc{r})],
\label{eq:nonres}\\
  &&\Ur(\vc{r}) = \frac1{\varepsilon_0}\sum_k
 n(\omega_{kn})\vc{d}_{nk}\sprod
 \re\{\dGamma_{\omega_{kn}}(\vc{r})\}\sprod \vc{d}_{kn}
\label{eq:res}.
\end{eqnarray}
Here, $\xi_j = j 2\pi\kB T/\hbar$ are the Matsubara frequencies;
$n(\omega)=[\rme^{\hbar\omega/(\kB T)}-1]^{-1}$ is the mean thermal
photon number of radiation of frequency $\omega$ at temperature $T$;
$\bm{\alpha}_n$ is the atomic polarisability; and the sum in
Eq.~(\ref{eq:res}) runs over all states $|k\rangle$ to which there
exist atomic dipole transitions with dipole moment matrix elements
$\vc{d}_{nk}=\langle n|\hat{\vc{d}}|k\rangle$ and energies
$\hbar\omega_{kn}=E_k-E_n$. For polar molecules \cite{ellingsen09}
and Rydberg atoms \cite{crosse10} alike, only a few transitions to
neighboring states contribute noticeably to the potential.

The tensor
\be
  \dGamma_\omega(\vc{r})
  \equiv\frac{\omega^2}{c^2}
  \dyad{G}^{(1)}(\vc{r},\vc{r};\omega)
\label{eq:gamma}
\ee
is given in terms of the scattering part $\dyad{G}^{(1)}$ of the
classical Green tensor for the electromagnetic field. We shall show
that this is a frequency-independent constant in the nonretarded limit.
To this end it is convenient to use a Dyson equation to represent
$\dyad{G}^{(1)}$ as a perturbative expansion in $\chi/(1+\chi/3)$
where $\chi(\vc{r};\omega)=\varepsilon(\vc{r};\omega)-1$ is the
dielectric susceptibility. As shown in Ref.~\cite{golestanian09},
\begin{align}  
&\dyad{G}^{(1)}(\vc{r},\vc{r}';\omega) =\frac1{k^2}\sum_{n=0}^\infty
 \int d \vc{s}_1 \cdots\int d\vc{s}_n
 \frac{\chi(\vc{s}_1;\omega)}{1\!+\!\third\chi(\vc{s}_1;\omega)}
 \notag \\
&\times\cdots
 \frac{\chi(\vc{s}_n;\omega)}{1\!+\!\third\chi(\vc{s}_n;\omega)}\,
 \dyad{A}_{\bd{\rho}_{r1}}\!\cdot\!\dyad{A}_{\bd{\rho}_{12}}
\!\cdots\!\dyad{A}_{\bd{\rho}_{nr'}}\label{eq:expansion}
\end{align}
with $k=\omega/c$, $\bd{\rho}_{r1}=\vc{s}_1-\vc{r}$,
$\bd{\rho}_{12}=\vc{s}_2-\vc{s}_1$ etc.\ and
\begin{align}
\dyad{A}_{\bd{\rho}}=
&-\frac{\rme^{\rmi k\rho}}{4\pi\rho^3}
\{[1-\rmi k\rho-k^2\rho^2]\dyad{I}\notag\\&
 -[3-3\rmi k\rho
-k^2\rho^2]\vc{e}_\bd{\rho}\vc{e}_\bd{\rho}\}\label{A}
\end{align}
($\dyad{I}$: unit matrix; $\vc{e}_\bd{\rho}=\bd{\rho}/\rho$). We have
assumed that all bodies are non-magnetic and isotropic, and that
$\vc{r}$ and $\vc{r}'$ lie outside the dielectric bodies, where
$\chi=0$.

In the nonretarded limit $|k\tilde{z}|\ll 1$, where $\tilde{z}$ is
the largest relevant particle--body distance, Eq.~(\ref{A}) reduces to
\be\label{eq:windependent}
\dyad{A}_\bd{\rho}
\simeq -(\dyad{I}-3\vc{e}_\bd{\rho}\vc{e}_\bd{\rho})/(4\pi\rho^3)
+ \mathcal{O}(k^2)\;.
\ee
We further assume that the perturbative parameter is
frequency-independent in some low-frequency range,
\be\label{eq:windepedent2}
\frac{\chi(\vc{r};\omega')}{1\!+\!\third\chi(\vc{r};\omega')}
\simeq\frac{\chi(\vc{r};0)}{1\!+\!\third\chi(\vc{r};0)}
\quad\mbox{for }|\omega'|\le\omega,
\ee
which is a good approximation for metals. With these two assumptions,
the expansion~(\ref{eq:expansion}) for the tensor~(\ref{eq:gamma})
can be approximated by
\begin{align}
 &\dGamma_{\omega'}(\vc{r})\simeq\dGamma_0(\vc{r})
 =\sum_{n=0}^\infty\frac1{(4\pi)^{n+1}}
\int d \vc{s}_1 \cdots\int d\vc{s}_n\notag \\
&\times\frac{\chi(\vc{s}_1;0)}{1\!+\!\third\chi(\vc{s}_1;0)}
\cdots\frac{\chi(\vc{s}_n;0)}
{1\!+\!\third\chi(\vc{s}_n;0)}\,
\dyad{A}_{\bd{\rho}_{r1}}\!\cdot\!\dyad{A}_{\bd{\rho}_{12}}
\!\cdots\!\dyad{A}_{\bd{\rho}_{nr}}
\label{eq:expansion2}
\end{align}
for $|\omega'|\le\omega$ together with Eq.~(\ref{eq:windependent}),
meaning $\dGamma_\omega(\vc{r})$ is \emph{independent} of
frequency in this range.

We apply this result to the CP potentials~(\ref{eq:nonres}) and
(\ref{eq:res}) in the nonretarded regime $|\omega_{kn}\tilde{z}/c|\ll
1$. The dominant contribution to the Matsubara sum is for
$\xi_j\lesssim|\omega_{kn}|\ll c/\tilde{z}$, so we can set
$\dGamma_{\rmi\xi_j}\simeq\dGamma_0$. Performing the sum according to
$\sum_{j=0}^{\infty\prime}\bm{\alpha}_n(\rmi\xi_j)%
=\sum_k[n(\omega_{kn})+\half]\vc{d}_{nk}\vc{d}_{kn}/(\kB T)$,
we obtain
\be\label{eq:nresapprox}
 \Unr(\vc{r})\simeq -\frac1{\varepsilon_0}\sum_k
 [n(\omega_{kn})+\half]\vc{d}_{nk}\sprod\dGamma_0(\vc{r})\sprod
 \vc{d}_{kn}.
\ee
Noting that $\chi(\vc{r};0)$ is real due to the Schwarz reflection
principle, Eq.~(\ref{eq:expansion2}) shows that
$\re(\dGamma_{\omega_{kn}})\simeq \re(\dGamma_0)=\dGamma_0$ in
Eq.~\eqref{eq:res}.

The total nonretarded potential hence reads
\be\label{eq:finalU}
  U_n(\vc{r})\simeq
 -\frac1{2\varepsilon_0}\sum_k\vc{d}_{nk}\sprod\dGamma_0
(\vc{r})\sprod \vc{d}_{kn}.
\ee
This remarkably simple expression is manifestly independent of both
temperature and transition frequencies $\omega_{kn}$. One can
explicitly verify that it coincides with the zero-temperature result
using $2\pi\kB T\sum_{j=0}^{\infty\prime}\to \hbar\int_0^\infty d\xi$
in Eq.~\eqref{eq:nonres} and $n_T(\omega)\to-\Theta(-\omega_{kn})$ in
Eq.~\eqref{eq:res} [$\Theta$: unit step function]. The
temperature-independent CP potential is governed by the electrostatic
Green tensor $\dGamma_0$. It is entirely due to the dipole
fluctuations of the particle. For geometries where image-charge
methods apply, the potential coincides with the quantum average of the
interaction energy of the particle's dipole with the image dipoles
inside the bodies \cite{lennard32}. The potential depends on the
atomic internal state only via the dipole fluctuations
$\sum_k\vc{d}_{nk}\vc{d}_{kn}%
=\langle\hat{\vc{d}}\hat{\vc{d}}\rangle_n$. There is hence no
qualitative difference between forces on ground-state or excited
atoms. In particular, the sign of the force is the same in both cases.

Note that the temperature-invariance holds for a particle in a chosen
energy eigenstate or any temperature-independent incoherent
superposition state. When the particle is in a \emph{thermal}
superposition of energy eigenstates, the CP potential acquires a weak
temperature-dependence as discussed in Ref.~\cite{ellingsen10}. In
addition, the internal dynamics of a particle initially prepared in
an energy eigenstate does of course depend on the ambient temperature.

%%%%%%%%%%%%% SECTION %%%%%%%%%%%%%%%%%%%%%%%%%%%%%%%%%%%%%%%%%%%%%%%%

\paragraph*{Criteria for temperature-independence.}

The $T$-inva\-riant form~\eqref{eq:finalU} of the CP potential relies
on three conditions. In the following, we explicitly state these
criteria and give the leading corrections due to slight violations.

(A) The relevant particle--body separations must be sufficiently
nonretarded such that
\be
\frac{\kB  T}{\hbar|\omega_{kn}|} 
\left(\frac{\tilde{z}\omega_{kn}}{c}\right)^2
 \ll 1. \label{ineq:ret}
\ee
Corrections due to retardation arise from corrections to
Eq.~(\ref{eq:windependent}). Provided that the
approximation~(\ref{eq:windepedent2}) holds, one has
$\dGamma_{\omega}\simeq\dGamma_0+\omega^2\dGamma''_0/2$ where
the primes indicate $\omega$-derivatives and
$\chi(\vc{r};\omega)\to\chi(\vc{r};0)$ is understood. We substitute
this Taylor expansion into Eqs.~(\ref{eq:nonres}) and (\ref{eq:res}).
In the geometric high-temperature regime
$\kB T\gg\hbar c/\tilde{z}\gg\hbar|\omega_{kn}|$, the nonresonant
potential does not contribute, because the $j=0$ Matsubara term
vanishes and all terms $j\geq 1$ are exponentially small. The
retardation correction to the CP potential is hence dominated by its
resonant contribution and it reads
\begin{equation}
\Delta U^\mathrm{retard.}_n
= \frac{\kB T}{2\hbar\varepsilon_0}
\sum_k\omega_{kn}\vc{d}_{nk}\sprod\dGamma_0''(\vc{r})\sprod
\vc{d}_{kn}.
\label{eq:total2}
\end{equation}
where $\dGamma''_0(\vc{r})$ is real. It can be reduced by choosing
particles with larger transition wavelengths or going to smaller
particle--body separations.

(B) The reflectivity of the bodies must be sufficiently frequency
independent such that
\be
\frac{\kB T}{\hbar|\omega_{kn}|}\,\|\dGamma_0(\vc{r})\|^{-1}
\|\re\dGamma_{\omega_{kn}}^\mathrm{nret}(\vc{r})-\dGamma_0(\vc{r})\|
\ll 1; \label{ineq:cond}
\ee
Here, $\dGamma_{\omega_{kn}}^\mathrm{nret}(\vc{r})$ is given by
Eq.~(\ref{eq:expansion}) with $\dyad{A}$ approximated by
Eq.~\eqref{eq:windependent}, but without assuming
Eq.~\eqref{eq:windepedent2}. Combining Eqs.~(\ref{eq:nresapprox}) and
(\ref{eq:res}), the reflectivity correction for
$\kB T\gg\hbar|\omega_{kn}|$ reads
\begin{equation}
\Delta U^\mathrm{refl.}_n
=\frac{\kB T}{\hbar\varepsilon_0}\sum_k\frac{\vc{d}_{nk}\sprod
[\re\dGamma^\mathrm{nret}_{\omega_{kn}}(\vc{r})-\dGamma_0(\vc{r})]
\sprod\vc{d}_{kn}}{\omega_{kn}}\,.
\label{eq:noncancel1}
\end{equation}

As examples of simple convex, planar and concave geometries, consider
an atom at distance $z$ from the surface of a sphere (radius $R\ll
z$) \cite{sambale09}, plate \cite{ellingsen10} or inside a spherical
cavity (radius $R=z$) \cite{sambale07}. In these cases, one finds
\begin{equation}
\frac{\Delta U_n^\mathrm{retard.}}{U_n(T\!=\!0)}
=c_\mathrm{retard.}\,\frac{\kB T}{\hbar\omega_{kn}}
 \biggl(\frac{z\omega_{kn}}{c}\biggr)^2
\end{equation}
with $c_\mathrm{retard.}=-1/3$ (sphere), $c_\mathrm{retard.}=0$
(plate) and $c_\mathrm{retard.}=3/5$ (cavity). The reflectivity
corrections read
\begin{equation}
\frac{\Delta U_n^\mathrm{refl.}}{U_n(T\!=\!0)}
=-c_\mathrm{refl.}\,\frac{\kB T}{\hbar\omega_{kn}}\,
 \frac{z|\omega_{kn}|}{c}\,
 \mathrm{Re}\frac{i}{\sqrt{\varepsilon(\omega_{kn})}}
\end{equation}
with $c_\mathrm{refl.}=R/z$ (sphere), $c_\mathrm{refl.}=6$ (plate) and
$c_\mathrm{refl.}=3$ (cavity). For a conductor with plasma frequency
$\omega_P$ and relaxation constant $\gamma$, the Drude model
$\varepsilon(\omega)=1-\omega_P^2/[\omega(\omega+i\gamma)]$ leads to
($\omega,\gamma\ll\omega_P$, $\omega$ real)
\begin{equation}
\re\bigl[\rmi/\sqrt{\varepsilon(\omega)}\bigr]
\simeq \omega_P^{-1}
 \left[\half(\sqrt{\omega^2+\gamma^2}+|\omega|)
 |\omega|\right]^\frac1{2}.
\end{equation}

The retardation and reflectivity corrections are additive to linear
order. More generally, the leading correction for high temperatures
can be obtained by combining Eq.~\eqref{eq:res} with the dominant
$j=0$ contribution from Eq.~\eqref{eq:nonres},
\begin{equation}
\Delta U_n(\vc{r},T)=
\frac{\kB T}{\hbar\varepsilon_0}\sum_k\frac{\vc{d}_{nk}\sprod
\Delta\dGamma_\omega(\vc{r})
\sprod\vc{d}_{kn}}{\omega_{kn}}\,+ \mathcal{O}(T^0),
\label{eq:noncancel}
\end{equation}
with $\Delta\dGamma_\omega(\vc{r})%
=\re\dGamma_{\omega_{kn}}(\vc{r})-\dGamma_0(\vc{r})$; it contains
both retardation and reflectivity corrections.

As will be demonstrated below, the reflectivity correction is
geometry-dependent. To wit, Eq.~\eqref{eq:noncancel} distinguishes
three classes of geometries where 
$\Delta \dGamma_\omega\buildrel{\omega\to 0}\over\propto%
|\omega|^\eta$ with (i) $\eta>1$, (ii) $\eta=1$ or (iii) $\eta<1$.
In case (i), the reflectivity correction can be made arbitrarily small
for sufficiently small $\omega$, and the $T$-dependence can be made
arbitrarily small. For case (iii), the slope of the linear correction
term $\propto T$ reaches a minimum as $\omega$ decreases; it
\emph{increases} when further reducing $\omega$ because the reduction
in the Green tensor difference is overcompensated by the growth of the
photon number.

(C) The CP potential must be dominated by transitions for which (A)
and (B) hold. If the conditions hold for a transition $n\to k$, but
not for $n\to l$, then one must have 
\be
\left(\frac{|\vc{d}_{nl}|}{|\vc{d}_{nk}|}\right)
\left(\frac{\kB T}{\hbar|\omega_{ln}|}\right)\ll 1,
\ee
ensuring that the linear $T$-corrections from retarded but suppressed
atomic transitions do not contribute. Care must be taken in the case
where a transition resonates with a cavity of quality-factor
$\mathcal{Q}$. Then we even require that
$\mathcal{Q}(|\vc{d}_{nl}|/|\vc{d}_{nk}|)(\kB T/\hbar|\omega_{ln}|)
\ll 1$.

%%%%%%%%%%%%% SECTION %%%%%%%%%%%%%%%%%%%%%%%%%%%%%%%%%%%%%%%%%%%%%%%%

\paragraph*{Numerical examples.}

We now present two illustrative examples of non-planar geometries with
curvature radii on the order of the particle--body separation. For
simplicity, we consider isotropic particles for which
\be\label{eq:trG}
 \vc{d}_{nk}\sprod\dGamma_\omega(\vc{r})\sprod \vc{d}_{kn}
 = \third|\vc{d}_{nk}|^2\Tr \dGamma_\omega(\vc{r}).
\ee

\begin{figure}[tb]
  \includegraphics[width=3in]{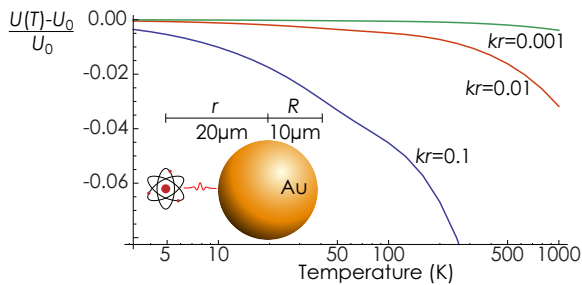}
  \caption{Comparison of the exact thermal CP potential $U(T)$
  of a ground-state two-level particle (transition energy $\hbar c k$)
  outside a gold sphere with the $T$-independent result $U_0$ from
  Eq.\ \eqref{eq:finalU}.}
  \label{fig:spheregraph}
\end{figure}

Consider first the case of a particle at distance $r$ from the center
of a gold sphere of radius $R$. The Green tensor for this geometry is
given in Ref.~\cite{buhmann04}. In the nonretarded limit $kr,kR\ll 1$
and for a perfectly conducting sphere $|\varepsilon|\to \infty$,
Eqs.\ \eqref{eq:finalU} and \eqref{eq:trG} lead to
\be\label{eq:perfectSphere}
U_n(\vc{r})\simeq-
  \frac{R^3(6r^4-3r^2R^2+R^4)}{
24\pi\varepsilon_0
r^4(r^2-R^2)^3}
\sum_k|\vc{d}_{nk}|^2.
\ee
This $T$-independent potential is in agreement with the
zero-temperature result found in Ref.~\cite{taddei10}. In
Fig.~\ref{fig:spheregraph}, we demonstrate the accuracy of Eq.\
\eqref{eq:perfectSphere} over a large range of temperatures for
particles of different transition frequencies outside a gold sphere.
%%%%%%%%%%%%% FIGURE %%%%%%%%%%%%%%%%%%%%%%%%%%%%%%%%%%%%%%%%%%%%%%%%%
%%%%%%%%%%%%%%%%%%%%%%%%%%%%%%%%%%%%%%%%%%%%%%%%%%%%%%%%%%%%%%%%%%%%%%
The sphere is described by the Drude model with $\omega_P=9$eV and
$\gamma=35$meV. As Fig.~\ref{fig:spheregraph} shows, the correction
due to finite temperature is about $1\%$ at room temperature when
$kr=0.01$ and decreases rapidly as $kr\to 0$, thus exemplifying case
(i) with $\eta=3/2$. The temperature correction due to retardation (A)
is the dominant correction, and the reflectivity correction (B) is
considerably smaller.

\begin{figure}[tb]
  \includegraphics[width=3.1in]{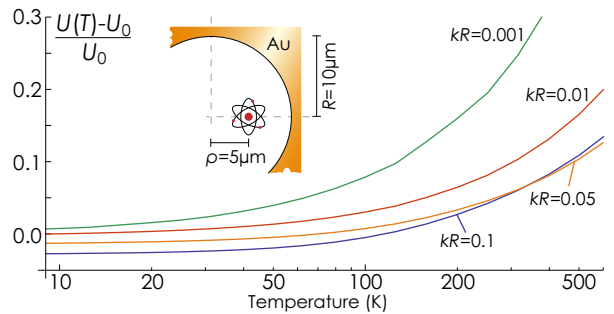}
  \caption{Same as Fig.~\ref{fig:spheregraph}, but for
  a particle inside  a gold cylindrical cavity.}
  \label{fig:cylgraph}
\end{figure}

Consider next a particle inside a cylindrical cavity of radius $R$
whose Green tensor can be found in Ref.~\cite{ellingsen09b}. In the
nonretarded limit and for a perfectly conducting sphere, it leads to
the potential
\begin{align}
U_n(\vc{r})\simeq&-
\int_0^\infty\frac{dq}{6\pi^2\varepsilon_0}
\msum\frac{K_m(qR)}{I_m(qR)}\left\{\left[(m/\rho)^2+q^2\right]\right.
\nonumber\\
  &\left.\times
I^2_m(q\rho)+q^2I_m^{\prime
2}(q\rho)\right\}
\sum_k|\vc{d}_{nk}|^2
\label{gammacylPC}
\end{align}
($K_m,I_m$: modified Bessel functions). Its reliability is illustrated
in Fig.~\ref{fig:cylgraph}. In contrast to the previous example,
the reflectivity correction (B) dominates; it may be shown that
except very close to the walls one has case (iii) with $\eta\approx
1/2$. Hence as the particle's transition frequency is decreased, the
temperature corrections reach a minimum. Note that this problem does
not arise when suppressing these corrections by reducing the spatial
dimensions of the setup. The temperature dependence is nonetheless
quite modest; for our parameters the linear $T$-correction is at the
$5$\% level at room temperature for $kR\sim 0.01-0.05$.

%%%%%%%%%%%%% FIGURE %%%%%%%%%%%%%%%%%%%%%%%%%%%%%%%%%%%%%%%%%%%%%%%%%
%%%%%%%%%%%%%%%%%%%%%%%%%%%%%%%%%%%%%%%%%%%%%%%%%%%%%%%%%%%%%%%%%%%%%%

%%%%%%%%%%%%% SECTION %%%%%%%%%%%%%%%%%%%%%%%%%%%%%%%%%%%%%%%%%%%%%%%%

\paragraph*{Conclusions.}

We have shown that the CP potential of a particle prepared in an
energy eigenstate at nonretarded distance from a conducting body may
be approximated by an expression which is independent of both
temperature and particle transition frequency, regardless of the
geometry of the body. In this regime, the interaction is governed by
the transition dipole matrix elements of the particle and the
electrostatic Green tensor. There is no qualitative difference between
the potential acting on atoms in ground or excited states. For
instance, this implies that repulsive CP forces that have been
predicted for ground-state atoms in certain geometries \cite{levin10}
may be more easily verified using excited particles in the same
geometry.

While the fact that dispersion interactions are of electrostatic
nature in the nonretarded regime is well known, the remarkable
cancellation of temperature dependent contributions only occur once
both nonresonant and resonant contributions are fully accounted for.
Each of these force components, which arise from physically distinct
phenomena, individually depend strongly on temperature and transition
frequency.

The two numerical examples illustrate the reliability of the
temperature-independent potential. For a particle next to a metallic
sphere, the temperature dependence vanishes for ever smaller values of
the retardation parameter $r\omega/c$, whereas for a particle in a
cylindrical cavity, the $T$-dependent corrections reach a minimum
beyond which they cannot be further reduced. The latter effect seems
to be a peculiarity of the cylindrical geometry.

This work was supported by the UK Engineering and Physical Sciences
Research Council (EPSRC).

%%%%%%%%%%%%%%%%%%%%%%%%%%%%%%%%%%%%%%%%%%%%%%%%%%%%%%%%%%%%%%%%%%%

\end{document}